\documentclass[twocolumn,letterpaper,prb,superscriptaddress,showpacs]{revtex4-1}
\usepackage{graphicx}
\usepackage{subfigure}
\usepackage{amsmath}
\usepackage{mathrsfs}
\usepackage{comment}
\usepackage{color}


\begin{document}

\title{Pressure-tuned Frustration of Magnetic Coupling in 
   Elemental Europium
        }
\author{Shu-Ting Pi}
%\email{spi@ucdavis.edu}
\affiliation{Department of Physics, University of California Davis, 
Davis, California 95616 USA}
% \author{Wei-Ting Chiu}
% \affiliation{Department of Physics, University of California Davis, One Shields Avenue,
% Davis, California 95616 USA}
\author{Sergey Y. Savrasov}
\affiliation{Department of Physics, University of California Davis, 
Davis, California 95616 USA}
\author{Warren E. Pickett}
\email{wepickett@ucdavis.edu}
\affiliation{Department of Physics, University of California Davis,
Davis, California 95616 USA}
\date{\today}

\begin{abstract}
Applying linear response and the magnetic force theorem in correlated
density functional theory, the inter-sublattice exchange constants
of antiferromagnetic Eu are calculated and found to vanish near the pressure of
P$_c$=82 GPa, just where magnetic order is observed experimentally to be
lost. The Eu $4f^7$ moment remains unchanged at high pressure, again in
agreement with spectroscopic measurements, leaving the picture of
perfect frustration of interatomic Ruderman-Kittel-Kasuya-Yoshida 
couplings in a broad metallic background, leaving a state of electrons
strongly exchange coupled to arbitrarily oriented, possibly quasistatic 
local moments. This strongly frustrated state gives way to 
superconductivity at T$_c$=1.7K, observed experimentally.
These phenomena, and free
energy considerations related to correlations, suggest an
unusual phase of matter that is discussed within the scenarios of the
Doniach Kondo lattice phase diagram, the metallic spin glass class, 
and itinerant spin liquid or spin gas systems.
\end{abstract}

\maketitle

% \section{Introduction}
The behavior of local moments and their ordering as some external
parameter (volume, electron density, magnetic field) varies lies at 
the root of several paradigmatic
phenomena, viz. the Kondo effect, heavy fermion
superconductivity, spin liquid, and spin glass phases. 
The $4f$ shell in lanthanides ($Ln$) has provided a unique platform for
the study of several of these issues. Ce and Yb compounds, with their
$4f$ level near the Fermi energy, show  
$4f$-conduction electron coupling that can be tuned across the Doniach
critical point from antiferromagnet (AFM) to Kondo lattice
at ambient pressure. Reduction in volume is needed to 
drive other lanthanides into exotic phases. 
% The theory of magnetic coupling in $Ln$ and actinide metals has been
% advancing from several viewpoints,\cite{Pi2014,Pi2014b,Locht2016,Tapia2017} 
% but puzzling behavior remains to be understood. 

Experimentally, 
a study by  Jackson {\it et al.}\cite{Jackson2005}
of six $Ln$ metals with pressure tuned in the 5-12 GPa range
indicated a linear decrease in the magnetic ordering temperature 
T$_m$ roughly in proportion to the de Gennes factor of
the $4f$ ion. However, higher pressures bring more complex behavior 
due to structural transitions and band structure changes.
In the lanthanides ($Ln$) Tb, Dy, and Nd, T$_m$ varies as much as
150K through pressure ranges up to 1.5 mbar,\cite{Lim2015b,Lim2015,Song2017}
often non-monotonically. 
\begin{comment}
It has been demonstrated that T$_m$ in $Ln$ varies strongly and often
non-monotonically with pressure:
for Tb, T$_m$ varies non-monotonically between 50K and room temperature (RT) in the
15-140 GPa range;\cite{Lim2015b} Dy has T$_m$ above RT at 157 GPa;\cite{Lim2015}
 T$_m$ for Nd increases to 180K in the 40-90 GPa range\cite{Song2017} before falling back to 50K.  
\end{comment}
In Eu, however, after non-monotonic behavior in T$_m(P)$ due to structural 
transitions,\cite{Bi2011,Bi2016}
in the $Pnma$ structure that exists in a range around 80 GPa, T$_m$ falls to 11K 
at P$_c$=82 GPa whereupon magnetic order is replaced with superconductivity (SC)
up to 3K.\cite{Debessai2009} 
%    Yb displays related behavior, as we discuss below. 
% While elemental Eu and Yb are known to display deviations from trends in $Ln$ behavior,
% both have been found to display additional behavior that
% becomes a challenge to theory, as we describe.

Advances in modeling exchange coupling in $Ln$
metal\cite{Pi2014,Pi2014b,Locht2016,Tapia2017} 
have dealt with {\it ordering}.  This first order disappearance 
of order represents an {\it avoidance} of the antiferromagnetic (AFM) 
quantum critical point that is actively studied in weak AFMs.\cite{Knebel2006}
This paramagnetic (PM) phase, 
with its superconducting (SC) ground state in the midst of disordered spin,
 may provide a platform for learning more about $Ln$ magnetic interactions,
and perhaps more general issues about neighboring phases near a QCP,
possibly including a spin liquid or spin gas phase coexisting with SC. 
 
This behavior can be compared with that of Yb. Under pressure, 
Yb undergoes a valence
transition,\cite{Syassen1982,Fuse2004,Dallera2006,Ylvisaker2009} 
from divalent $f^{14}$ to 
somewhere near trivalent $f^{13}$ through a
continuous evolution through intermediate valence and emergence of a local moment, a
crossover that has been simulated successfully by dynamical mean field
calculations\cite{Ylvisaker2009} up to 40 GPa. 
Recently Song and Schilling have reported\cite{Song2018}
that Yb, notwithstanding its $f^{13}$ local moments, becomes superconducting in 
the 1.4-4.5 K range at 80 GPa and above.
This behavior has parallels with, but distinctions
to, that of Eu, to which we return to in the discussion.

The PM phase above P$_c$ is unusual in having large spins on a dense 
periodic lattice interacting 
via RKKY Heisenberg exchange (the spins are isotropic) yet they do not order, a
signature of a type of frustration that is not apparent.
Following the classification
of Sachdev and Read,\cite{Sachdev1996} we refer to this as the metallic
quantum paramagnet (MQPM) phase.
Beyond the question of (dis)ordering, there is the perplexing issue of 
superconductivity in a metal with disordered strong local moments.
A simple scenario would be that Eu would be driven through a 
valence transition to the
non-magnetic $f^6$ $J$=0 configuration, in which case there is no magnetic impediment to superconducting pairing, viz. the isovalent rare earth metal Y becomes an impressive
superconductor under pressure, with T$_c$ up to 17 K.\cite{Hamlin2006}
We find that Eu, unlike Yb, in not near a change in valence up to 100 GPa or more.

%Historically, 
While early studies suggested a valence transition below 80 GPa,\cite{Rohler}
%  and some interpretations of xray and
% spectroscopic data have been interpreted to indicate that the formal valence 
% is reduced significantly from $\nu$=3 for
%pressures in the few tens of GPa range.\cite{Rohler} 
more recent x-ray absorption data
confirms that Eu retains its $f^7$ moment even in the SC phase above 
82 GPa,\cite{Bi2012,Song2018} in agreement with our calculations.
The SC phase is then of an exotic type in which pairing occurs within a dense
lattice of large but disordered and uncompensated moments.  These questions
have led us to perform systematic calculations
of the electronic structure and magnetic coupling of Eu at pressures 
up to the 100 GPa range.

\begin{figure}[tbp]
\centering
\includegraphics[width=1.0\columnwidth]{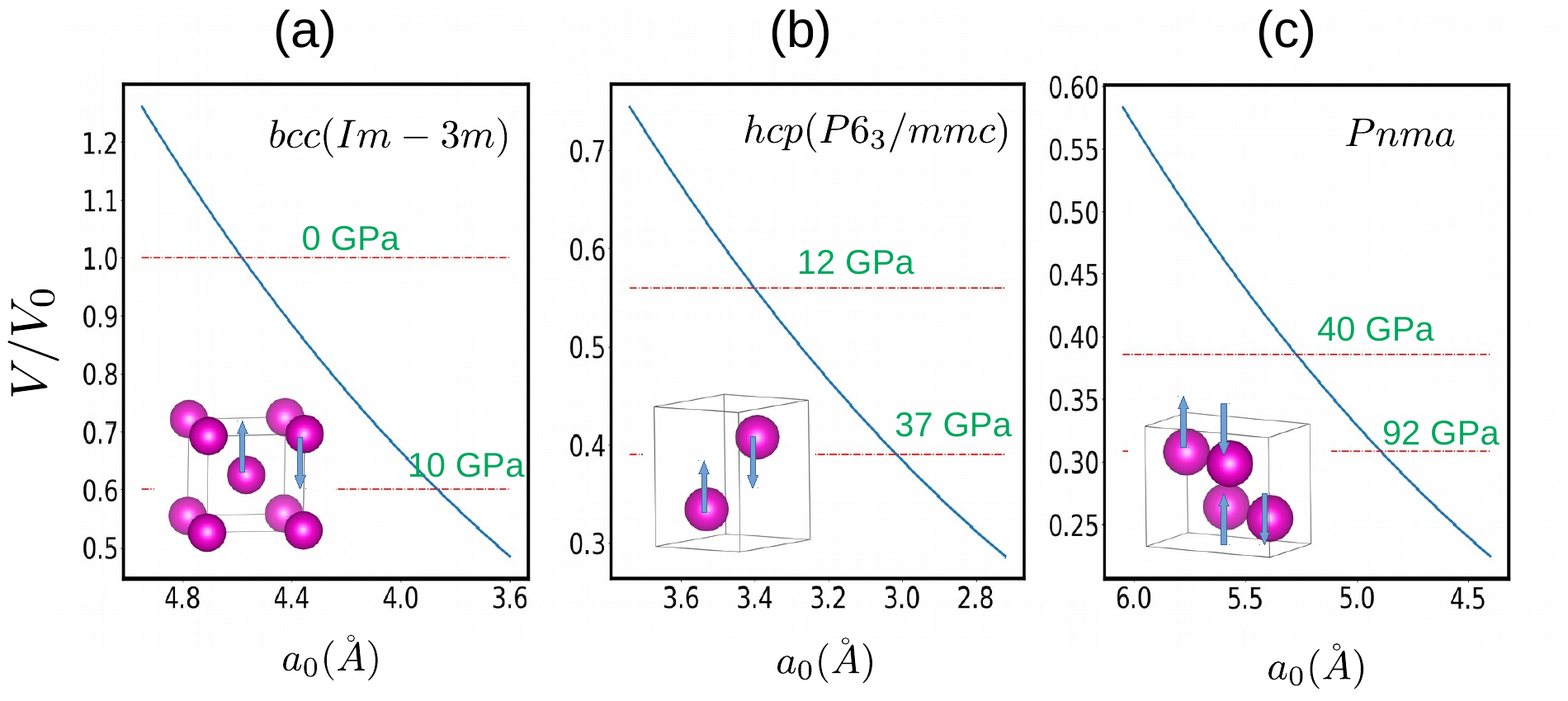}
\caption{(color online) The relative volume $V/V_{0}$ ratio as a 
function of the $a$ lattice constant  
(a) $bcc$ (b) $hcp$ (c) $Pnma$. The lattice structures were obtained from 
experiments. Red dash lines indicate the $V/V_{0}$ ratio of pressures 
where the structures are confirmed stable in experiments, as given
by Bi {\it et al.}\cite{Bi2011}}
\end{figure}

Like all $Ln$ metals,
Eu displays structural transformations with increasing pressure. Structural
information is provided in the Supplemental Material (SM).
%The bcc phase extends 
%from 0-12 GPa. The common bcc-to-hcp transition for $Ln$ metals occurs at 12 GPa, 
%with single phase hcp in the 12-18 GPa range. The 18-66 GPa regime is more complex,
%with mixed-phases reported.\cite{Husband2012} 
% reported that the mixed phases are mostly hcp+$C2/c$
% from 18 to 40 GPa, followed by $Pnma+C2/c$ from 40 to 66 GPa. 
%Above 66 GPa, a single phase $Pnma$ regime is reached, which covers the 
%pressure range of interest here. We will simplify and consider 
%the structures as 
%three stages: bcc, 0-12 GPa; hcp, 12-40 GPa; $Pnma$, $>$ 40 GPa. 
% In Table I the
%experimental lattice structures at specific pressures are listed.
The volume decrease $V/V_{0}$ ratio ($V_0$ is the ambient volume) and the
regions of stability of the three phases\cite{Bi2011} 
are shown with the magnetic structures in Fig. 1.
The evidence is that Eu displays AFM order from ambient 
to $P_c = 82$ GPa. At $P_c$, magnetic order vanishes and superconductivity 
emerges with critical temperature T$_c$=1.7K, increasing with pressure up to 
2.8K at 142 GPa.\cite{Debessai2009}
As recently reported\cite{Bi2016} and as we confirm from calculation,
the large moment on Eu persists (the $f^7$ moment in Gd is 
calculated\cite{Yin2006} to persist to 500 GPa), making
the interplay between large but disordered moments and SC, and 
its dependence on pressure, unresolved issues. 

%%%%%%%%%%%%%%%%%%%%%%%%%%%%%%%%%%%%%%%%%%%%%%%%
\begin{comment}
\begin{table}[tbp]
\caption{Crystal structures of Eu metal in various pressures. $x,y,z$ are the Wyckoff
positions, and there are 1, 2 and 4 atoms per primitive
cell, respectively. Only in the bcc case is the unit cell doubled. }% 
\begin{ruledtabular}
\begin{tabular}{ccccccc}
P =  &  4 GPa  & 14 GPa & 75 GPa &
\\ \hline
Structure & $Im{\bar 3}m$ & $P6_3/mmc$ & $Pnma$ \\
  & $bcc$ & $hcp$ & orthorhombic \\
\hline
Lattice  & a=4.1961 & a=3.3501 & a=4.977  \\
constants &         & b=3.3501 & b=4.264  \\
 ($\AA$) &          & c=5.2962 & c=2.944  \\
\hline
x  & 0 & 1/3 & 0.325  \\
y  & 0 & 2/3 & 1/4 \\
z  & 0 & 1/4 & 0.029 \\
\end{tabular}
\end{ruledtabular}
\end{table}
\end{comment}
%%%%%%%%%%%%%%%%%%%%%%%%%%%%%%%%%%%%%%%%%%

%  \subsection{Methods}
Our density functional theory (DFT) calculations employ the full 
potential linearized-muffin-tin-orbital method 
(LMTO).\cite{Savrasov1992} The local spin density approximation (LSDA) 
with Hubbard U correction (LSDA+U) on the localized $4f$ shell orbitals 
of Eu. A reasonable 
value is $U$ = 6-7 eV at ambient pressure; at high pressure we investigate
smaller values of $U$. Note that we use LSDA+U rather than LDA+U because 
the spin-density
mediated intra-atomic $f-d$ Hund's coupling that polarizes the conduction electrons  
is important to include and assess.

%Rather than attempt to calculate conduction electron mediated RKKY coupling
%out to several neighbors, 
Interatomic RKKY exchange constants are known to extend out to dozens 
of neighboring shells in Eu.\cite{Kunes2003,Turek2003}
Instead we focus on the AFM sublattice exchange constants,
which are linear combinations of interatomic exchange constants out to 
arbitrary distance.
An effective and efficient method 
is to use linear response theory and the magnetic force theorem.\cite{AIL} 
Consider the electronic Kohn-Sham Hamiltonian $H_{KS}=T+V_{0}+V_{sp}$, 
where $T$ is kinetic energy,
$V_{0}$ is the spin-independent potential, and $V_{sp}$ is the spin-dependent
potential including the contribution from $U$. We write 
%\begin{eqnarray}
$V_{sp}=\vec{\sigma} \cdot \textbf{B},~~
\textbf{B} = [v_{xc,\uparrow}(r)-v_{xc,\downarrow}(r)] \hat{B}$, 
%\label{effectiveB}
%\end{eqnarray}
where $\vec{\sigma}$
is the electron spin vector of Pauli matrices.  
\textbf{B} appears as an effective Zeeman
field arising from the spin-dependent exchange-correlation potential $v_{xc}$. 

If one rotates the moment on AFM sublattices $\tau$, $\tau^{'}$ in unit cells 
\textbf{R}, \textbf{R}$^{'}$ by infinitesimal angles 
$\delta$\textbf{$\theta$}$_{\tau R}$, $\delta$\textbf{$\theta$}$_{\tau^{'} R^{'}}$
respectively, 
the second order energy difference is related to the exchange constants by
\begin{align}
J_{\tau R \tau^{'} R^{'}}^{\alpha \beta} =& \frac{\delta^{2}E}{\partial 
\theta^{\alpha}_{\tau R}\partial \theta^{\beta}_{\tau^{'} R^{'}}}
=\sum_q J_{\tau\tau'}^{\alpha\beta}(q) e^{iq\cdot(R-R^{'})}, \\
%  \notag
%\end{align}
%The linear response expression in momentum space is
%\begin{align}
J_{\tau\tau'}^{\alpha\beta}(q)=&\sum_{kjj^{'}} 
B^{\tau\alpha}_{kj;k+q.j'} B^{\tau'\beta~*}_{kj;k+q.j'}
\frac{f_{kj}-f_{k+qj'}}{\epsilon_{kj}-\epsilon_{k+qj^{'}}}, \\
%  \langle kj| [\sigma 
%  \times \textbf{B}_{\tau}]_{\alpha}|k+qj^{'} \rangle \notag \\
%  & \times \langle k+qj^{'}| [\sigma 
 B^{\tau\alpha}_{kj;k+q.j'} =&\langle kj| [\sigma
\times \textbf{B}_{\tau}]_{\alpha}|k+qj^{'} \rangle.
\label{RKKY}
\end{align}
Here $j,j^{'}$ are band indices, $\alpha,\beta$ are Cartesian coordinates,
$k,q$ are wave vectors, $f_{kj}$ is the Fermi function, $\epsilon_{kj}$ and 
$|kj\rangle$ are the LSDA+U energies and eigenstates.  
This method has been confirmed to work well 
in several transition metal oxides and rare earth compounds.%\cite{Successes} 
~A version extended
to systems with strong spin-orbit coupling and multipolar exchange interactions
was also proposed and applied successfully.\cite{Pi2014,Pi2014b}

%%% DISCUSSION %%%

The initial questions to address are the $4f$ occupation and the position
of the $4f$ levels with respect to the Fermi energy E$_F$. Technical details
are provided in the SM. 
%In Fig. S1 of the
%Supplemental Material (SM), the $4f$ occupations (all majority spin) and the total
%moment inside the atomic sphere are displayed for bcc, hcp, and $Pnma$ phases
%across their pressure range of stability. 
For all structures and pressures studied, the full $S=\frac{7}{2}$
$4f$ contribution persisted, with a conduction band ($5d$) contribution of
0.1-0.2 $\mu_B$ when spins were aligned. 
%   The moments decreased somewhat with 
%   reduced volume, reflecting increasing overlap with conducting states.
% the $4f^7$
% configuration persists and polarizes somewhat the conduction states.  
% Figure S2 of SM shows the band structures, for each phase at pressures 
% near the bottom
% and top pressures of their stability range. 
The $4f$ bands are centered near -5 eV, with the main change with 
pressure being that the $4f$ band
``width'' increases, primarily a crystal field increase rather than a
hopping amplitude increase. 
%The $4f$ configuration
%is stable and inert, remaining at -5 eV for pressure of interest here. 
For comparison, the $4f^7$ configuration of Gd has been calculated to 
remain stable to 500 GPa and above.\cite{Yin2006}
To indicate the magnitude of the exchange constants and provide connection
with future experiment,  the spin
wave spectrum for the ambient pressure bcc phase was calculated and is
provided in the SM.

\begin{figure}[tbp]
\centering
\includegraphics[width=0.85\columnwidth]{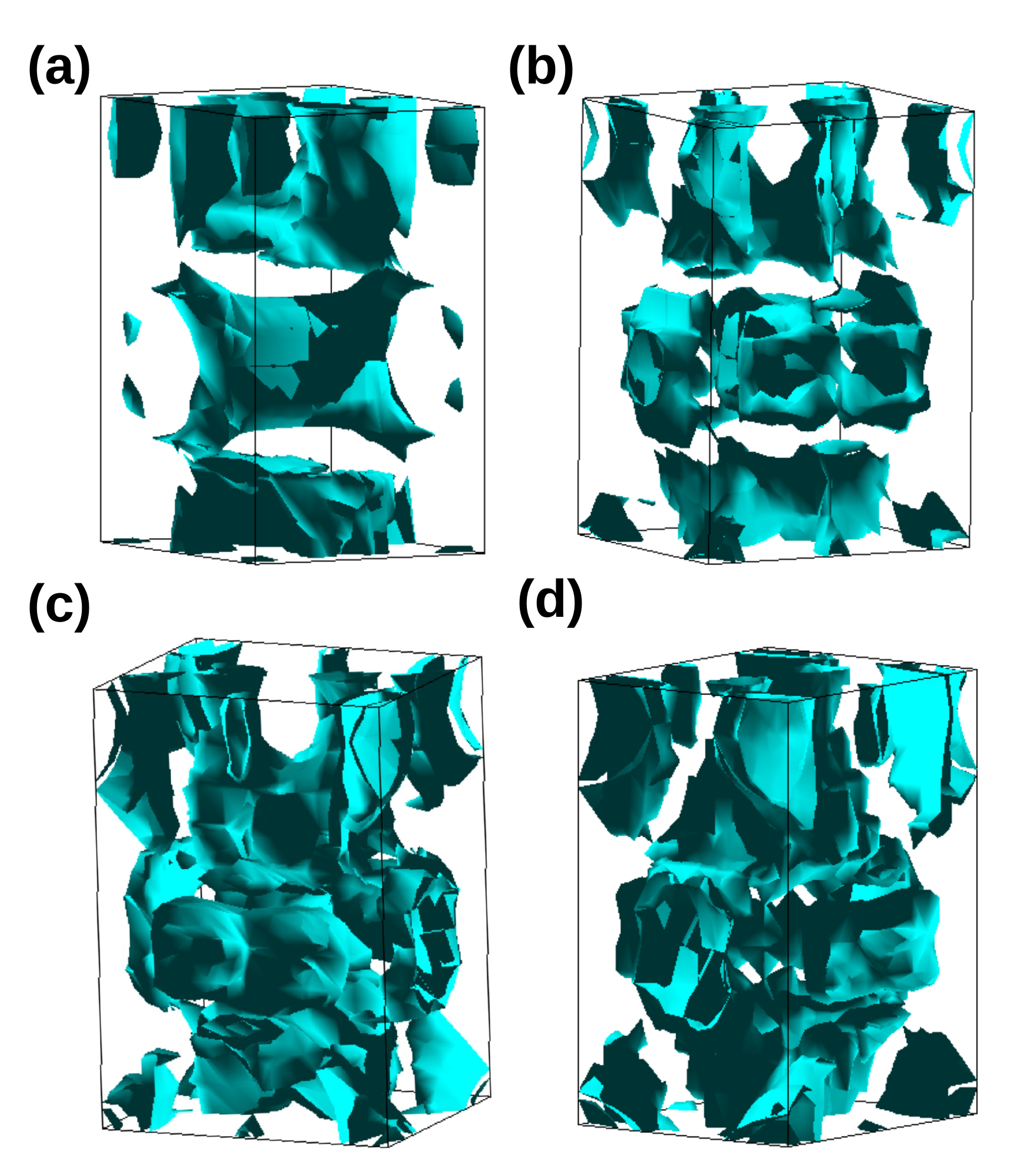}
\caption{(color) The Fermi surfaces of Eu at the relative volumes, with respect
to that at 75 GPa) of +6\%, 0, -4\%, and -10\%. At all volumes the surfaces
are large and multisheeted, varying through changes in topology
with the only effect being the decrease in exchange constants and hence
the ordering temperature, which vanishes around
82 GPa.
 }
\label{FS}
\end{figure}

%The spinwave velocity
%is in the $20-30$ meV~\AA~range with maximum energy $\hbar\omega_q^{max}$
%=12-16 meV, depending on the value of the Hubbard repulsion $U$.
%Experimental measurements would help in
%pinning down the effective $U$ value in Eu metal at low pressure.

\begin{figure}[tbp]
\centering
\includegraphics[width=1.0\columnwidth]{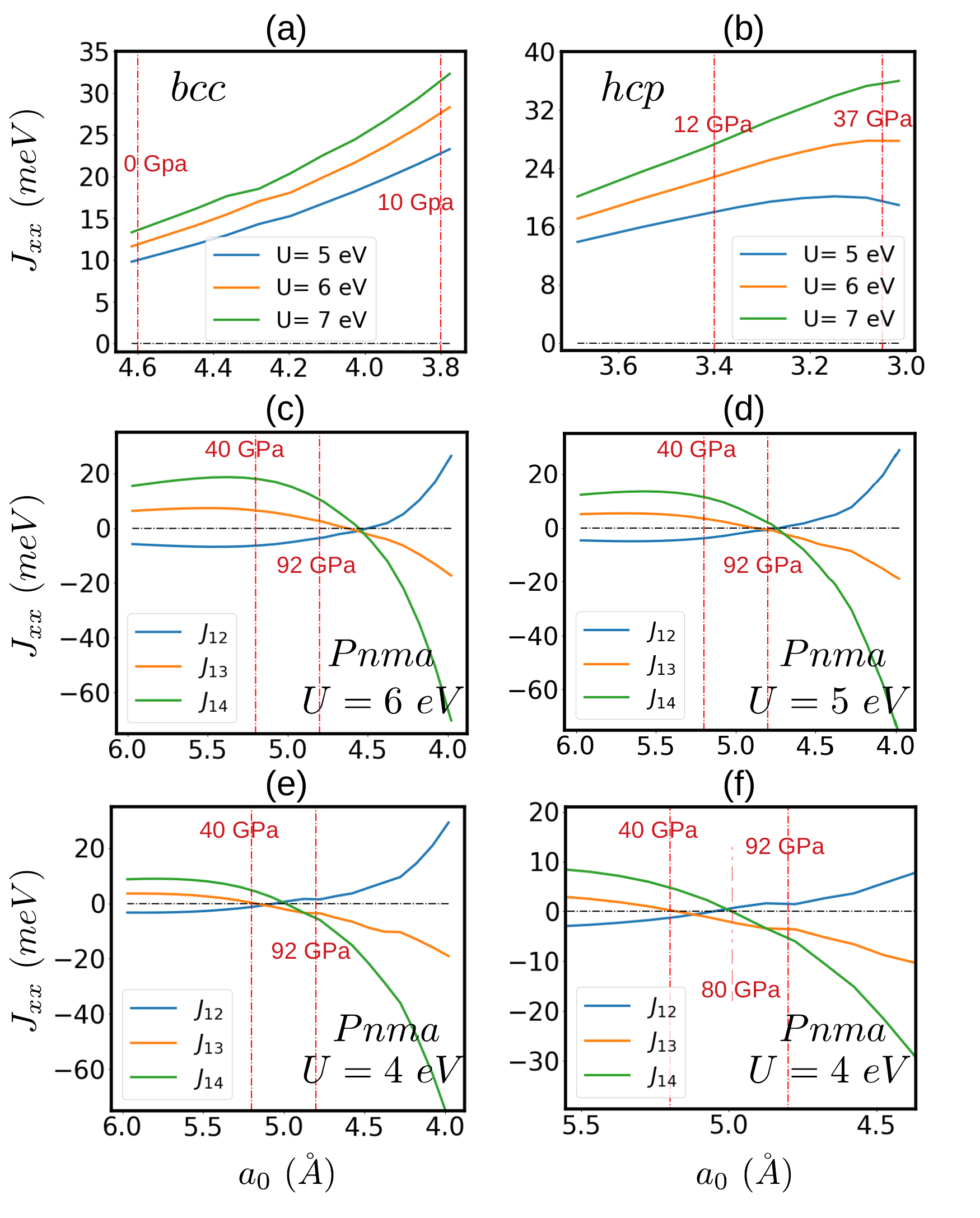}
\caption{(Color online) The exchange constants $J_{\tau\tau'}(q=0)$ for various 
pressures, as labeled. Panels (a) and (b) show bcc and hcp, respectively. 
The others are for the $Pnma$ phase at higher pressure: (c) U=6eV, (d) U=5eV, 
(e) U=4eV. 
Panel (f) focuses more closely on the sign change region in (e), showing that
the zero crossings lie close to 80 GPa. Red dashed lines represent experimental 
pressures at the displayed lattice constants. }
\label{exchange}
\end{figure}

Magnetic coupling of the $Ln$ metals in general and Eu in particular, with their
non-overlapping local moments within an itinerant electron sea, 
is due to the conduction electron mediated 
RKKY exchange mechanism described above. 
Throughout the pressure range studied, and in particular in the regime 
where magnetic order vanishes,
the Fermi surface is large and multisheeted but evolving, as pictured in Fig.~\ref{FS}. 
Large sheets are separating in the vicinity of $P_c$, but this change in Fermi surface 
topology does not lead to significant van Hove singularities in the density of states
nor to identifiable structure in $J_{ij}$ versus pressure.
While spiral magnetic order is
commonly identified with nesting of sheets of Fermi surface, stable AFM order
while the Fermi surface evolves argues against any nesting origin of ordering.

Equations (2-4) were used to calculate the sublattice exchange constants 
$J_{\tau\tau'}$
based on the AFM ordered state. 
%  each $J_{\tau\tau'}$ is a linear combination
%  of inter{\it atomic} RKKY exchange couplings. 
With two up and two down spins, symmetry reduces
the number of constants to three, viz. up1-up1, up1-down1, up1-down2, denoted
below by $J_{12}, J_{13}, J_{14}$ respectively.
%Since $4f^7$ ions are spherical and the
%electronic structure is three dimensional, the exchange
%%constant tensor is effectively diagonal in Cartesian coordinates.
The RKKY expression includes momentum-conserving
virtual excitations, with those near the Fermi level having larger weight.
In the $q\rightarrow 0$ limit, inter-sublattice exchange
constants contain distinct intraband and interband terms (for general
$q$ there is no distinction). % The former is the zone sum of a Fermi
% level density of states factor $|\vec v_k|^{-1}$ modulated by matrix elements. 
The energy denominator makes exchange coupling somewhat sensitive
to Fermi surface nesting, and several examples of incommensurate (often
spiral) order in lanthanides have been
traced back to Fermi surface calipers. The interband contribution 
%  may still be important but,
%  being unrelated to the FS, it 
will be continuous and more slowly varying than the intraband contribution.
The calculated $J_{\tau\tau'}(q=0)$ couplings
versus pressure are shown in Fig.~\ref{exchange} for $U$=5, 6, 7 eV.  
For bcc and hcp Eu, Fig.~\ref{exchange}(a),(b) respectively, the single
sublattice coupling is FM for $J_{12}$ and AFM in sign for the other two,
and each increases monotonically in magnitude over the range of interest.
% Larger $U$ enhances the coupling; AFM order is very stable in these two phases.  

The behavior in the high pressure $Pnma$ structure is different. 
In Fig. 4(c)-(f)  
$J_{12}$, $J_{13}$, and $J_{14}$ at $\vec q$=0 are shown, with increasing pressure
and for $U$=4, 5, 6 eV. $U$ affects primarily the magnitude, not changing the 
behavior as volume is reduced. The trend with increasing pressure is 
for {\it all three sublattice couplings} to decrease in magnitude and 
pass through zero nearly
simultaneously, signaling a collapse of the spinwave spectrum and frustration of
{\it sublattice coupling} rather than frustration of magnetic order. 
This trend is independent of the value of $U$; the collapse of coupling
-- the incipient QCP -- corresponds to the experimental observation of loss
of order best with $U=4.5$ eV. The collapse occurs at somewhat lower pressure
as $U$ is decreased. $U$ is expected 
to decrease under pressure from the $U=6-7$ eV value that is realistic at ambient
pressure.
% As a linear combination of RKKY couplings, the (simultaneous) 
% vanishing of sublattice exchange couplings indicates
% a cancellation of positive and negative couplings from various Eu-Eu neighbors, 
% that is, frustration of sublattice coupling.  
%It is not feasible to evaluate the $J_{\tau\tau'}$ on a fine enough
%$\vec q$ to invert Eq.~\ref{RKKY} and obtain small, long range 
%exchange constants, but they might not provide further understanding anyway. 
Note that the curves in Fig.~\ref{exchange}(c-f) become unphysical beyond
$P_c$. 
%  the underlying AFM order, which can still be obtained in the DFT+U
%  calculation, has become unstable, analogous to the phenomenon of unstable phonons.  
It is not unusual in highly frustrated magnets to encounter a range of
exchange couplings for which AFM order vanishes.

Eu thus provides a contrast to the Fe-pnictides where impact of magnetic 
interactions on the phase diagram has been actively studied. Our methods applied
to Fe pnictides led to (1) effective short-range coupling, 
and (2) AFM order that vanishes
due to first neighbor ($J_1$) and second neighbor ($J_2$) coupling as $J_1/2J_2$
approaches unity.\cite{Yin2008,Han2009}. Such a $J_1-J_2$ model near
frustration, with spins damped
by conduction electrons, was proposed by Wu {\it et al.}\cite{Si2008,Si2016} to account
for the quantum critical point versus isoelectronic As$\rightarrow$P doping in
BaFe$_2$As$_2$. Recently Sapkota {\it et al.}\cite{Sapkota2017} reported
near-perfect $J_1-J_2$ frustration in an itinerant metallic system,
square lattice CaCo$_{1.86}$As$_2$ tuned (naturally) by Co vacancies.
Frustration in Eu is in a stoichiometric lattice with local moments and
RKKY interactions, so the
mechanism of frustration -- volume evolution of many exchange constants --
is distinct. 

%  \section{Discussion}
{\it Discussion.}
Both experiment and our calculations concur that Eu retains its $f^7$ local moment
without valence change,\cite{Bi2016}
and magnetic coupling vanishes at $P_c$. 
Evidently the evolution of the electronic structure plays a 
critical role by inducing a AFM-MQPM transition. That the three independent
couplings vanish together at $P\approx P_c=$82 GPa suggests that the 
Kondo coupling between spin and conduction electrons dominates RKKY
coupling\cite{Isaev2013} and has decreased dramatically with pressure.
We have calculated the hybridization function\cite{Han2008} 
and determined that this 
is not the case.

\begin{figure}[tbp]
\centering
\includegraphics[width=1.0\columnwidth]{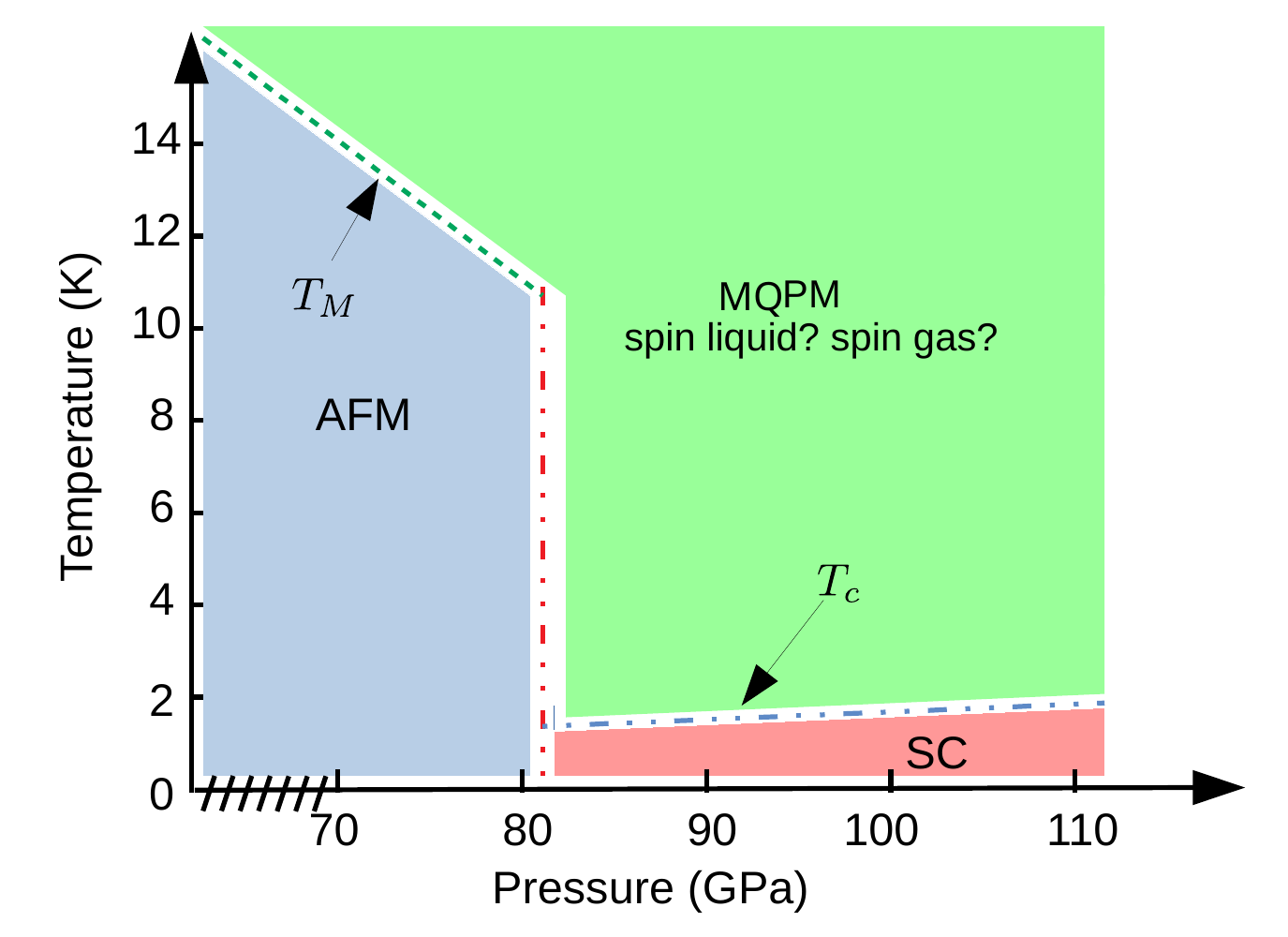}
\caption{(color online) Schematic depiction of the experimental phase
diagram of Eu under pressure, showing the first order transition
at P$_c$=82 GPa. Phases are: MQPM, metallic quantum paramagnetic; AFM: antiferromagnetic;
SC, superconducting. }
\label{phasediagram}
\end{figure}

A schematic phase diagram based on experimental data is
presented in Fig.~\ref{phasediagram}. Magnetic order decreasing to 11K vanishes
at $P_c$, and superconducting electronic order emerges as the ground state of this
MQPM.  
%whose driving force should be considered.
This ground state presents a potentially new phase: 
a superconducting condensate in the midst of large 
disordered moments (not compensated by Kondo coupling) below T$_c$=1.7K. 
The character of the transition from                       
AFM  to an MQPM phase at $P_c$ is related to the question
of magnetic correlations in the MQPM phase.
In the free energy 
%\begin{eqnarray}
$F(P,T)=E_{DFT}(V(P)) + PV_{DFT}(P) +E_m(P,T) -TS_m(P,T) $
%\end{eqnarray}
the first two are available from DFT calculations and are linear 
in P  and T-independent at low T since (1) no structural change 
occurs,\cite{Bi2011} and (2) the magnetic moments remain, only the 
order vanishes.\cite{Bi2016} 
The electronic entropy $\frac{\pi^2}{3}k_B T N(E_F)$
$\sim\frac{\pi^2}{3}k_BT/W$, where the bandwidth is $W=\sim 2-4$ eV,
is orders of magnitude smaller than the magnetic terms 
and has not been displayed.  

The magnetic contributions, $E_m$ and
$S_m$ from spinwaves in the AFM phase, or spin disorder in the
PM phase, must account for the small free energy change across the transition.
The difference in entropy between ordered and uncorrelated moments
at high temperature is S$^{\infty}$ =
$k_B ln(2S+1) = 3k_B ln 2$ for $S=\frac{7}{2}$.
A rough (factor of two) estimate of the entropy of the ordered phase can be
taken from the linear spin-wave expression S$_{AFM}=\beta(P)$T$^3$;
$\beta$ is P-dependent because it depends on the exchange couplings 
$\{J_{\tau\tau'}\}$.
The entropy just above T$_N$ is roughly $S^{\infty}$/2, a
common value for AFMs.  Equating these at T$_N$,
one obtains the change across the transition as temperature is lowered
for $P<P_c$
\begin{eqnarray}
\Delta [TS_m(P,T)]   %   &=& TS(P,T)|^{T>T_N}_{T<T_N} \nonumber \\
    \approx \frac{1}{2} S^{\infty} T [1 - \frac{T^3}{T_N(P)^3}],
\end{eqnarray}
which is smooth and small across the magnetic transition but becomes
sizable at somewhat lower temperature. 

However,
supposing uncorrelated moments for $P>P_c$, the change in entropy across
$P_c$ has the same form: the increase in entropy contributes to the
loss of magnetic order above $P_c$, with a finite jump for $T<T_N(P_c)$=11K
but vanishing at $(P_c,T_N)$, giving no driving force for a first order
transition {\it at this point} in the phase diagram.

The magnetic energy $E_m$ of thermally excited spinwaves, $E_m(P,T) =
\int d\omega D(\omega,P) n(\omega/T)$ in terms of the spinwave density of
states $D$ and the Bose occupation factor $n(\omega/T)$, is replaced above $P_c$
with contributions depending on the degree of magnetic correlation among
the disordered spins. 
Total lack of correlation is unrealistic, in fact considerable short-range
correlation must survive to leave only a small change in the free energy at
$P_c$. The result: the necessary small change in free energy across 
$P_c$ implies strong correlation between the moments in the MQPM phase.
Such a magnetic subsystem may exhibit behavior characteristic of a spin
liquid\cite{Savary2017} or that of a spin glass.\cite{Mydosh2015} 
YMn$_2$ and CaCo$_{1.86}$As$_2$ both are magnetic metals that have 
been discussed as spin liquids,\cite{Nakamura1997, Sapkota2017} but unlike
Eu they are understood in terms of frustrating short-range interactions.

In closing, we comment on the unconventional electronic state in the SC phase.
The scenario that has emerged is that of
superconducting pairs co-existing with a spin glass or spin liquid magnetic 
system, presumed classical given the large value of the moments. With negligible
quantum fluctuation and the temperature being low
compared to other scales, one has pairing in the midst of  
quasistatic spins.
%     (although virtual excitations may contribute to pairing). 
Superconductivity
in the context of spin glasses has been discussed, for example by Galitski
and Larkin,\cite{Galitski2002} and an example proposed by Davidov
{\it et al.},\cite{Davidov1977} however spin glasses are nearly always treated
in the dilute impurity limit where positional disorder is a central issue,
whereas the spins in Eu are dense and periodic.
Our calculated exchange splitting of the Eu $d$ bands for 
ferromagnetic alignment indicates a local
on-site $f-d$ Hund's exchange strength of 0.75 eV, corresponding to a FM Kondo
coupling of $K$=0.75/($\frac{7}{2}\times \frac{1}{2}) \sim$0.4 eV.
This strong coupling
suggests comparable spin-disorder broadening of the
conduction bands, hence washing out of the Fermi surface. Spin-disorder
is normally destructive of pairing, unless the mechanism actually
proceeds through, and depends on, the dynamic spin system. 
Such pairing, if it is
responsible, lies in a different regime in Eu than for
the cuprates, Fe-based superconductors, and heavy fermions, 
where magnetic fluctuations
of small moments are intimately mixed into the conduction states.
Yb at high pressure, as discussed in the introduction, presents a
SC phase that may possess similarities to that of Eu.

\begin{comment}
Elemental Yb provides a related high pressure phase. 
Though Yb is non-magnetic $f^{14}$
at ambient pressure, it undergoes
a valence crossover under pressure, becoming strongly mixed valent
by 40 GPa,\cite{Ylvisaker2009}, and recent x-ray absorption near-edge
spectra indicates it is still mixed valent (with majority $f^{13}$)
to 125 GPa.\cite{Song2018}
At 82 GPa superconducting T$_c$=
1.4K is observed, and rises with additional pressure.\cite{Song2018} 
Similarly to Eu,
this is a superconducting state in which strong disordered, localized
$4f$ moments coexist with paired conduction states, although the two materials
arrive at this state from different paths. It seems possible
that superconductivity in Eu and Yb may require a new mechanism for
pairing and long-range coherence, and may involve coexistence of superconductivity
with a spin liquid phase. 
\end{comment}

%  \section
%{\it Acknowledgments.}
We thank Wei-Ting Chiu, R. Kremer, R. R. P. Singh, J. Oitmaa, and 
J. S. Schilling for helpful 
communications on this project.
Support from DOE NNSA grant DE-NA0002908 (S.T.P.) and
NSF grant DMR-1607139 (W.E.P.) are gratefully acknowledged. 
For computational facilities, we used the
National Energy Research Scientific Computing Center (NERSC), a DOE
Office of Science User Facility supported by DOE under Contract No.
DE-AC02-05CH11231, as well as an in-house cluster at UC Davis.

\bibliographystyle{unsrt}

\end{document}